\documentclass[aps,preprint,groupedaddress]{revtex4-1}  

\usepackage{graphicx}  
\usepackage{dcolumn}   
\usepackage{bm}        
\usepackage{amssymb}   
\usepackage[english]{babel}

\begin{document}


\title{Engineered arrays of NV color centers in diamond \\
based on implantation of CN$^{-}$ molecules through nanoapertures}
\date{\today}

\author{P.~Spinicelli$^{1}$}
\author{A.~Dr\'eau$^{1}$}
\author{L.~Rondin$^{1}$}
\author{F.~Silva$^{2}$}
\author{J.~Achard$^{2}$}
\author{S.~Xavier$^{3 }$}
\author{S.~Bansropun$^{3 }$}
\author{T.~Debuisschert$^{3}$}
\author{S.~Pezzagna$^{4}$}
\author{J.~Meijer$^{4}$}
\author{V.~Jacques$^{1}$}
\author{J.-F.~Roch$^{1}$}

\affiliation{$^{1}$Laboratoire de Photonique Quantique et Mol\'eculaire, ENS Cachan, UMR CNRS 8537, F-94235 Cachan cedex, France}
\affiliation{$^{2}$Laboratoire d'Ing\'enierie des Mat\'eriaux et des Hautes Pressions, UPR CNRS 1311, F-93430 Villetaneuse, France}
\affiliation{$^{3}$Thales Research and Technology, Campus Polytechnique,
F-91767 Palaiseau cedex,  France}
\affiliation{$^{4}$RUBION, Ruhr-Universit$\ddot{a}$t Bochum, D-44780 Bochum, Germany}

\begin{abstract}


We report a versatile method to engineer arrays of nitrogen-vacancy (NV) color centers in diamond at the nanoscale. The defects were produced in parallel by ion implantation through 80 nm diameter apertures patterned using electron beam lithography in a PMMA layer deposited on a diamond surface. The implantation was performed with CN$^-$ molecules which increased   the NV defect formation yield. 
This method could enable the realization of a solid-state coupled-spin array and could be used for  positioning   an optically active NV center on a photonic microstructure.

\end{abstract}

\maketitle 

\section{Introduction}

\indent The nitrogen-vacancy (NV) color center in diamond, consisting of  a substitutional nitrogen atom (N) associated to a vacancy (V) in an adjacent lattice site of the   crystalline matrix,  has found a wide range of applications in quantum information processing.
It is a robust luminescent center which, once isolated at the individual level \cite{Gruber_Science_1997}, can be used to implement  efficient single-photon quantum key distribution protocols \cite{Alleaume_NJP_2004}. Moreover, the  association of a spin structure in the ground level having a  long coherence time at room temperature 
with   spin dependent optical transitions 
allows for quantum state preparation by optical pumping 
and single-spin quantum state readout \cite{Jelezko_PRL_2004}.  
However, development of most of the envisioned applications require to position the NV center with  nanometer-scale accuracy. This spatial control could be  the basic  technology  for building a scalable quantum simulator based on  spins associated to an array of NV defects with magnetic dipolar coupling \cite{Neumann_NatPhys_2010}. The control of the position of a NV defect could also help to enhance the photon out-coupling  by  channeling  its luminescence toward a photonic structure, like a dielectric nanowire \cite{Babinec_NatureNano_2009} or a metallic antenna \cite{Greffet_PRL_2009}.

\indent  
A practical solution can consist in addressing NV centers in diamond nanocrystals 
\cite{Treussart_PhysB2006}. By manipulating the nanoparticle with a  tip-based system 
 \cite{Ampen-Lassen_OptExpress_2009,Vandersar_APL_2009}, it is then possible to place an hosted NV defect at a controlled location. This technique was used to couple the luminescence of a single NV center to   a photonic waveguide \cite{Benson_OptLett_2009} and could be scaled up to build arrays of NV defects.  
However, the   dipolar magnetic coupling between  two   NV centers is limited to a distance of a few tens of nanometers in the case of  millisecond spin coherence time reached in ultrapure single-crystal  diamond grown using the chemical vapor deposition (CVD) method \cite{Gopi_NatMat2009}. Due to an uncontrolled level of impurities and to parasitic surface effects, the spin coherence of NV centers in nanodiamonds is up to now limited to   much smaller values and  this bottom-up approach cannot be realistically envisioned, even  with touching  state-of-the-art luminescent nanodiamonds   
\cite{Tisler_ACSNano2009}. 
 
 \indent 
Ion-beam nitrogen implantation is a flexible technique to create individual NV centers in a   diamond sample \cite{Meijer_APL_2005,Rabeau_APL_2006} and two-qubit coupling was obtained between   implanted nitrogen atoms coming from the dissociation of a nitrogen molecule impinging on a diamond surface   
\cite{Neumann_NatPhys_2010,Gaebel_NaturePhysics_2006}. However, the resolution of   a nitrogen ion beam is limited   to a few hundreds of nanometers even with ion optics correction \cite{Meijer_NuclearMethods_2002}, thus preventing the scale-up of this clever recipe. Alternative techniques enabling the reliable placement of nitrogen impurities into a diamond substrate have therefore to be developed.  
Spatially resolved ion implantation with nanometer resolution can be achieved  by limiting the aperture of a low-energy ion beam with a hole made in a scanning probe  which will then define the implantation spot  
\cite{Meijer_AppliedPhysicsA_2008,Schenkel_JVacSciTechnol_2008} or by the  release  from a Paul-like ion trap in which the ions were   captured and cooled 
\cite{Schnitzler_PRL_2009}.  
 In this letter, we report a versatile technique consisting in ion implantation through an array of apertures  which was patterned in a   PMMA layer  initially deposited on top of the diamond substrate. Our   results demonstrate that this technique is well adapted to the parallel implantation of any impurity inside a large number of targeted spots, with a resolution defined by the diameter of the aperture.  

\section{Sample}

\begin{figure}[t] 
\centerline{\includegraphics[width=12cm]{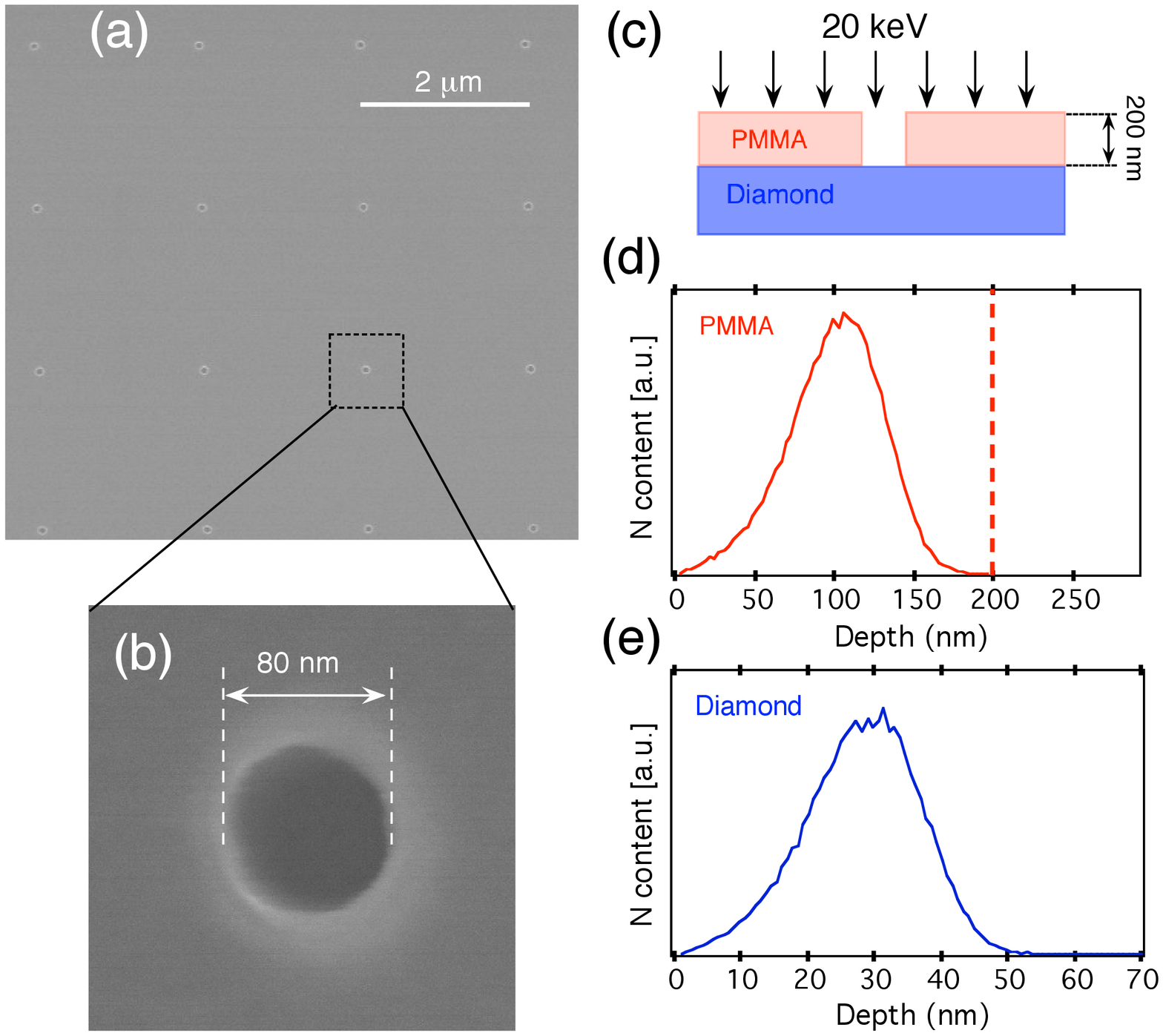}}
\caption{(a)-Scanning electron microscope image of the surface of the PMMA layer, displaying the array of apertures drilled using electron beam lithography. (b)-Enlarged image of the PMMA layer surface  zooming a single aperture. (c), (d), and (e)-Implantation scheme, with SRIM simulations corresponding to the 20~keV nitrogen irradiation of the PMMA layer and the unmasked diamond surface. According to the simulation shown in (d), the ions are fully stopped by the 200 nm thickness of the PMMA layer. }  
\label{Fig1}
\end{figure}

We used an ultra-pure CVD-grown single-crystal diamond with an intrinsic nitrogen content  much below 1~ppb~\cite{Tallaire_DRM_2006}, ensuring that each detected NV center can be faithfully attributed to an implanted nitrogen impurity~\cite{Naydenov_APL_2010}. 
After acid cleaning of the sample, 
a 200~nm thick layer of PMMA A4 was deposited on the diamond surface by spin-coating. Arrays of 80~nm diameter apertures were then patterned on the PMMA layer using a 80 kV-2.3 nA electron beam lithography equipment (Nanobeam-Ltd). The apertures were regularly spaced with a $2 \  \mu$m pitch array. After electron beam lithography, the mask was finally developed using MIBK/IPA. It resulted in an array of apertures with controlled spacing and diameter, as shown in Fig.~\ref{Fig1} (a) and (b).
This process, which simply relies on the deposition of a PMMA resist and patterning using electron beam lithography, prevents from any damage of the diamond substrate, as compared to other methods like plasma etching or carving  with a focused ion beam.

\section{Implantation}

Nitrogen was then implanted from a CN$^{-}$ molecular ion beam with 40 keV kinetic energy. When a  CN$^{-}$ molecule hit the surface, it broke in two parts. Taking into the mass selection in the ion beam, it then   became  equivalent    to a joint implantation of nitrogen  $^{14}$N  and carbon $^{12}$C  with respective energies   about 20 keV and 18 keV.  As shown in Fig~\ref{Fig1} (d) and (e), a SRIM simulation \cite{Ziegler_SRIM} indicates that the nitrogen atoms   will stop at an approximate depth of $30 \pm 10$  nm below the diamond surface, whereas they are stopped by the PMMA layer in the masked parts of the sample (see Fig.~\ref{Fig1} (c)). The evaluated straggling was about 9 nm, much smaller than the diameter of the aperture. Note that the simultaneous braking of the nitrogen and carbon atoms as they penetrate into the diamond will  increase the number of vacancies  created  close to the implanted nitrogen atoms, thus helping in the formation of the NV centers   
 \cite{Naydenov_APL_2010}.


\indent The toxic CN$^{-}$ molecules were {\it in situ} produced   with  a negative sputter ion source. Cs ions were used to sputter a target consisting of a mixture of BN and graphite powders 
 and reaction in the plasma generated above the target led to the creation of  CN$^{-}$ molecules  among other species. In this ion source which was initially developed for tandem accelerator, the negatively charged ions  were only produced  in their single charged state. This effect decreased the number of possible ions corresponding to a given energy-mass product, providing an   easy identification of a given ion which selection was realized  by a double focused   $90^{\circ}$ magnet. 
The sorted ions were then accelerated to an adjustable kinetic energy ranging from a few keV to hundreds of keV. The output of the accelerator is equipped with an implantation setup consisting of an   electrostatic scanning system  and a specifically designed chamber with secondary electron suppression~\cite{Ion_Channeling}.

\indent    After the implantation run corresponding to an applied dose of $10^{12}$ ions/cm$^2$ with a current of 600 nA,  the   PMMA layer was removed by acid treatment. The    sample was then annealed at $800^{\circ}$C for 2 hours to induce vacancy diffusion leading to the conversion  of the implanted nitrogen atoms into luminescent  NV color centers.

\section{Optical characterization of the implanted sample}

\begin{figure}[b] 
\centerline{\includegraphics[width=12cm]{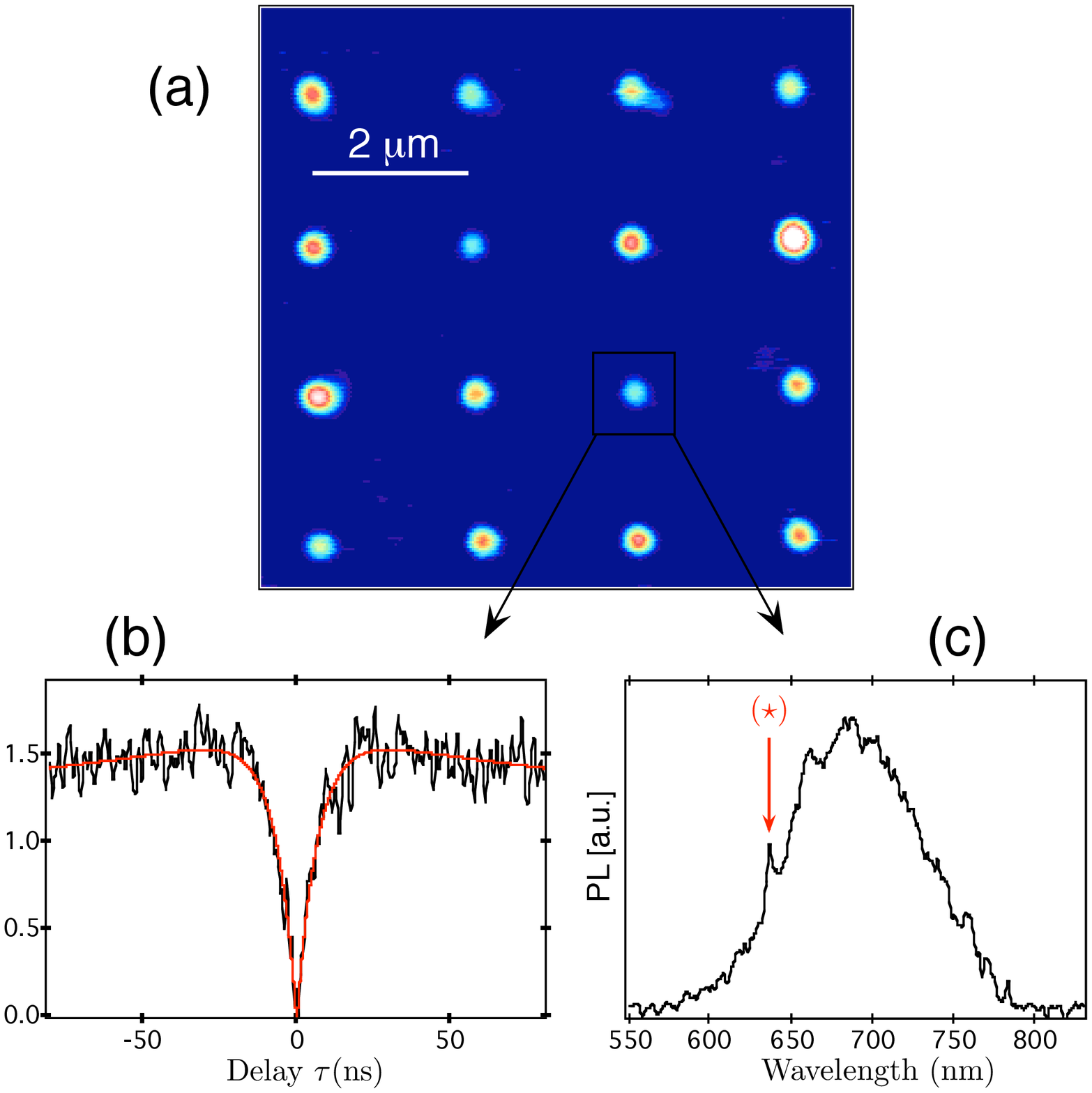}}
\caption{(a) PL raster scan of the sample, displaying an emission pattern matching the array of holes  drilled in the PMMA layer. (b) and (c)- Analysis realized for each spot, corresponding   to the determination of the second-order autocorrelation function associated to  the time delay $\tau$ between consecutive photon detections, and to the record  of the luminescence spectrum. The zero-phonon line ($\star$ symbol) at   637 nm   identified the center as negatively charged NV$^-$.  The signal over background associated to a single emitter, as observed in   the circled spot,  was near 4.    }
\label{Fig2}
\end{figure}

\indent The photoluminescence (PL) properties of the implanted and annealed sample were then studied using a  home-made scanning confocal microscope. A laser operating at 532~nm wavelength  was focused onto the sample through an oil immersion microscope objective. The PL was collected by the same objective, spectrally filtered from the remaining excitation laser light, and focused onto a $50 \ \mu$m diameter pinhole. The PL was finally directed   to avalanche photodiodes working in photon counting regime.
A typical PL raster scan of the sample is shown in Fig.~\ref{Fig2}~(a), displaying an array of photoluminescent spots   which matches the PMMA mask with its 
period of drilled holes. This observation clearly demontrates the efficiency of the implantation technique. \\

\indent
In order to infer the number of emitters associated to each luminescent spot, we performed a photon correlation measurement   using a Hanbury Brown and Twiss interferometer. 
Following the procedure of Ref.~\cite{Brouri_OptLett_2000} for   background correction and normalization to shotnoise level,   the histogram of the time delays $\tau$ between two consecutive photon detections is equivalent to the second-order autocorrelation function $g^{(2)} (\tau)$ of the luminescence intensity emitted by the analyzed spot.  
 Fig.~\ref{Fig2} (b) and (c) respectively show the photon correlation measurement and  the spectral analysis associated to the circled   spot in   Fig.~\ref{Fig2} (a).  The dip observed at zero delay in the   $g^{(2)} (\tau)$  function  is the signature of  a single emitter   located in the corresponding implanted area.
  Complementary spectral analysis showed that most of the observed emitters were   of    the negatively-charged state NV$^{-}$. Note that in our case of shallow implantation, the remaining defects of neutral charge state NV$^{0}$ could be efficiently transformed into their counterpart NV$^{-}$ by surface oxidation, as shown in Ref.~\cite{Santori_APL2010}.

 
\begin{figure}[h!] 
\centerline{\includegraphics[width=12cm]{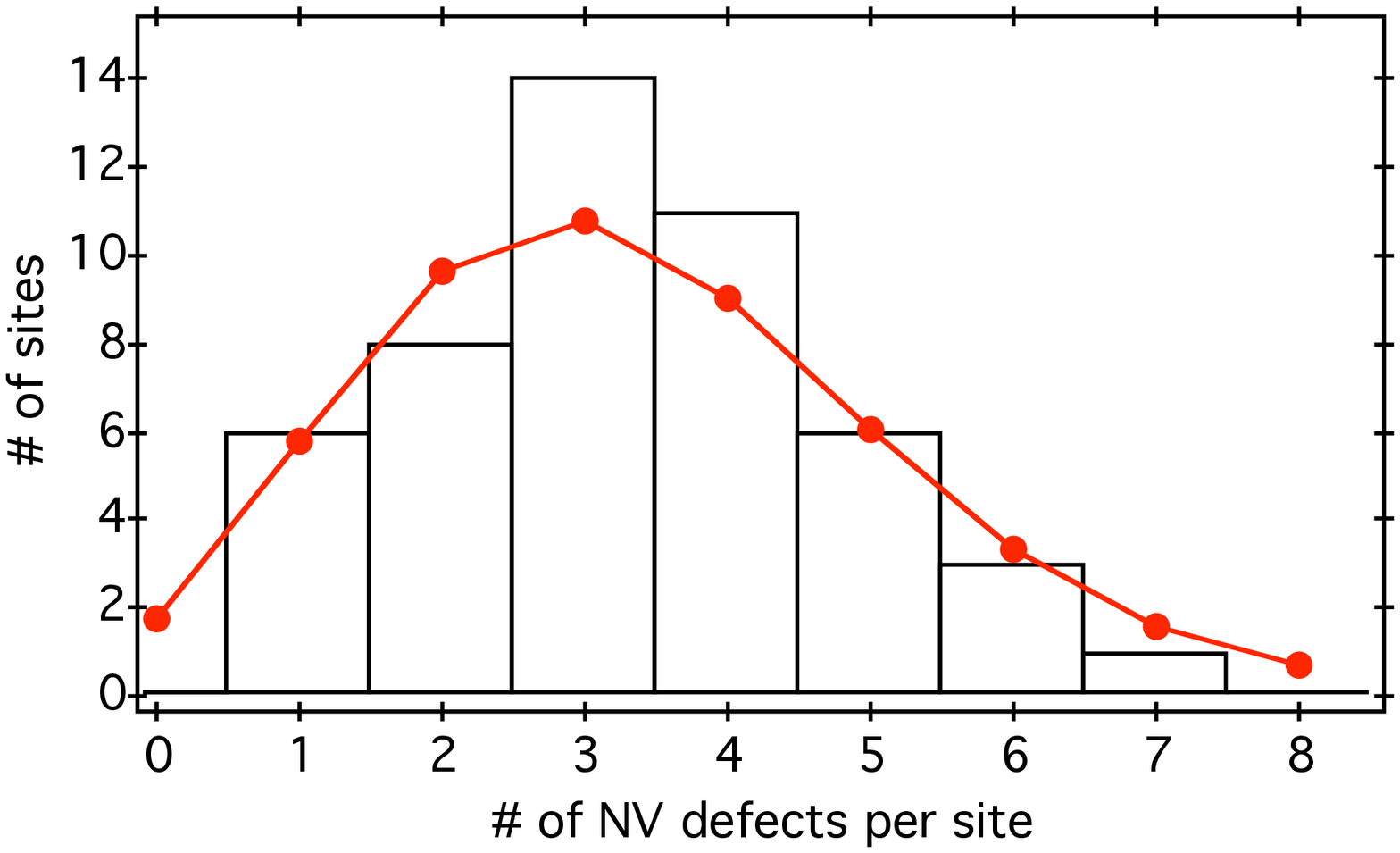}}
\caption{Statistics evaluated on 49 sites of the number of NV defects per implanted spot, and   fit by a Poisson distribution.}
\label{Fig3}
\end{figure}

 The number $N$ of emitting NV defects in the implanted spots was finally determined from  combined measurements of the depth of the zero-delay dip  in the  $g^{(2)} (\tau)$ function,  and of the luminescence intensity  compared to the mean level corresponding to a single NV emitter. The corresponding histogram  is shown in  Fig.~\ref{Fig3}. Since the number of implanted ions is random, we obtained a broad distribution which can be fitted by a Poissonian distribution with a mean value equal to $3.5$.   Since the irradiation dose 
 corresponds to approximately 50   nitrogen atoms per spot, the conversion yield of implanted nitrogen ions into NV defects is near 0.07. As expected from the joint implantation of carbon atoms associated to the the CN$^-$ molecular beam, this value is more than two times higher than the reported ones for low-energy nitrogen  implantation \cite{Rabeau_APL_2006,Pezzagna_NJP_2010}.
 


\section{Conclusion}

In conclusion, we showed that arrays of NV color centers can be reliably produced by implanting CN$^{-}$ molecules through the 80 nm diameter holes in a PMMA mask of 200 nm thickness which was deposited on top of a single-crystal diamond sample. The joint implantation of carbon increased the conversion of the implanted   nitrogen atoms into NV centers. The apertures and the thickness of the PMMA layer have to be optimized in order to improve the   resolution of the implanted impurities but fabrication of patterns below 10 nm have already been reported   in thinner PMMA layers
 \cite{Vieu_AppliedSurfaceScience_2000}. This flexible technique holds therefore strong promise for building scalable  quantum registers based on coupled spins.  


\section{Acknowledgements}

We are grateful to G\'eraldine  Dantelle for processing the diamond sample after the implantation run, and we thank Fedor Jelezko for helpful discussions. Work done at LPQM was partially sponsored by the DIAMAG project of Agence Nationale de la Recherche.
 During the writing of our manuscript, we have been aware of a related work reporting   nitrogen implantation   through an array of apertures in an electron beam lithography resist~\cite{Awschalom_Nanoletters_2010}.


\begin{thebibliography}{99}
\bibitem{Ion_Channeling} 
The implantation chamber is tilted  at $7^{\circ}$ in order to avoid ion beam channeling which could  induce an increase of the ion penetration depth  compared to the SRIM prediction. The    efficiency of the design   was controlled by an independent    SIMS measurement which gave a derivation below  $5\%$ in dose and depth for a silicon sample implanted  with 60 keV B$^{+}$ ions.
\end{thebibliography}
\end{document}